\documentclass[aps,prd,groupedaddress]{revtex4}
\usepackage{graphicx}
\usepackage{epstopdf}

\begin{document}

\title{Model Independent Form Factors for Spin Independent Neutralino-Nucleon Scattering from Elastic Electron Scattering Data}

\author{Gintaras D$\bar \textrm{u}$da}
\email{gkduda@creighton.edu}
\author{Ann Kemper}
\email{kemper@creighton.edu} \affiliation{Department of Physics,
Creighton University, 2500 California Plaza, Omaha, NE 68178, USA}
\author{Paolo Gondolo}
\email{paolo@physics.utah.edu} \affiliation{Department of Physics,
University of Utah, 115 S 1400 E Rm 201, Salt Lake City, Utah
84112-0830, USA}

\begin{abstract} Theoretical calculations of neutralino-nucleon interaction rates with
  various nuclei are of great interest to direct dark matter
  searches such as CDMS, EDELWEISS, ZEPLIN, and other experiments
  since they are used to establish upper bounds on the WIMP-proton
  cross section.
  These interaction rates and cross sections are generally computed with standard, one or two
  parameter model-dependent nuclear form factors, which
  may not exactly mirror the actual form factor for the particular
  nucleus in question.  As is well known, elastic electron scattering can allow for very precise
  determinations of nuclear form factors and hence nuclear charge densities
  for spherical or near-spherical nuclei. We use charge densities derived from elastic electron
  scattering data to calculate model independent, analytic form
  factors for various target nuclei important in dark matter
  searches, such as Si, Ge, S, Ca and others.  We have found that for
  nuclear recoils in the range of 1-100 keV significant
  differences in cross sections and rates exist when the model
  independent form factors are used: at 30 keV nuclear recoil the
  form factors squared differ by a factor of 1.06 for $^{28}$Si, 1.11 for $^{40}$Ca,
  1.27 for $^{70}$Ge, and 1.92 for $^{129}$Xe.  We show the effect of different form factors on the
  upper limit on the WIMP-proton cross section obtained with a hypothetical $^{70}$Ge detector during a
  100 kg-day effective exposure. Helm form factors with various parameter choices differ at most by 10--20\%
  from the best (Fourier Bessel) form factor, and can approach it to better than 1\% if the parameters are
  chosen to mimic the actual nuclear density.
\end{abstract}

\maketitle

\section{Introduction}

Neutralinos, or more generically weakly-interacting massive
particles (WIMPs), may be detected in the laboratory ``directly"
as they elastically collide with nuclei in a target, depositing
enough energy to give typical nuclear recoils of keV. See
\cite{1}-\cite{smith} for example for a review of calculating
detection rates of neutralino dark matter.

Direct detection experiments set limits on the WIMP-proton or
neutralino-proton cross section in the following manner.  First,
the spin-independent neutralino-nucleus elastic cross section with
a pointlike nucleus of $Z$ protons and $N$ neutrons can be written
as
\begin{equation}
\sigma^{SI}_{i} = \frac{\mu^2_{i}}{\pi}\left|Z G^p_s + (A-Z)
G^n_s\right|^2,
\end{equation}
where $G^p_s$ and $G^n_s$ are the scalar four-Fermion couplings of
a WIMP with point-like protons and neutrons \cite{gondolo96} and
$\mu_i=mM/(m+M)$ is the WIMP-nucleus reduced mass, with $m$ the
neutralino mass and $M$ the nucleus mass.  In most theoretical
models it is assumed that $G^p_s = G^n_s$, so that the cross
section scales with the square of the nucleus atomic number $A$.
Thus the spin independent WIMP-proton cross section can be written
as

\begin{equation}
\label{sigmasi}
\sigma^{SI}_i = \sigma_p A^2 \left(\frac{\mu_i}{\mu_p}\right)^2
\end{equation}

For a given detector, the expected number of recoil events with
recoil energy in the range ($E_1$,$E_2$) is the sum over the
nuclear species in the detector given by

\begin{equation}
N_{E_1-E_2} = \sum_i \int^{E_1}_{E_2} \frac{dR_i}{dE}
\mathcal{E}_i(E) dE \label{N},
\end{equation}

\noindent where $\mathcal{E}_i(E)$ is the effective exposure of each
nuclear species in the detector (introduced in \cite{compatibility with dama}), and $dR_i/dE$ is the expected
recoil rate per unit mass of species $i$ per unit nucleus recoil
energy and per unit time.  The effective exposure is given by

\begin{equation}
\label{efficiency} \mathcal{E}_i = \mathcal{M}_i T_i
\epsilon_i(E),
\end{equation}

\noindent where $T_i$ is the active time of the detector during
which a mass $\mathcal{M}_i$ of nuclear species $i$ is exposed to
the signal, and $\epsilon_i(E)$ is the counting efficiency for
nuclear recoils of energy $E$.  The differential rate $dR_i/dE$ is given by

\begin{equation}
\label{eq:2} \frac{dR_i}{dE} = \frac{\rho \sigma^{SI}_i \left|
F(q) \right|^2 }{2 m \mu^2_i} \int_{v>q/2\mu} \frac{f({\vec
v,t})}{v} d^3 v. \label{dR/dE}
\end{equation}
Here $E$ is the energy of the recoiling nucleus, $\rho$ is the
local halo WIMP density,  $f({\vec v,t})$ is the WIMP velocity
distribution function in the frame of the detector,
$\sigma^{SI}_i$ is the spin-independent WIMP-nucleus elastic cross
section off a pointlike nucleus, and $|F(q)|^2$, with
$q=\sqrt{2ME}$ (the recoil momentum), is a nuclear form factor.
The upper bound on the WIMP-proton cross section is computed by
comparing the expected number of events given by (\ref{sigmasi}-\ref{eq:2}) to the observationally
set limit on the number of events detected in the relevant energy
range(s).

The determination of an upper limit on the WIMP-proton cross
section depends of course on the shape of the signal in $dR/dE$,
which in turn depends on two factors: 1) the integral over the
WIMP velocity distribution, which contains the astrophysical
uncertainties and which we will not discuss here, and 2) the form
factor $|F(q)|^2$, which contains the nuclear physics
uncertainties and which we will examine. Notice that the upper
limit set by direct detection experiments on $\sigma_p$ is
inversely proportional to $|F(q)|^2$.  So a change of $|F(q)|^2$
by a factor of two implies an upper limit on $\sigma_p$ stronger
by a factor of two. It is thus crucial to understand and have a
correct value for the nuclear form factor for the relevant range
of momentum transfers.

The nuclear form factor, $F(q)$, is taken to be the Fourier
transform of a spherically symmetric ground state mass
distribution normalized so that $F(0)=1$:
\begin{equation}
F(q) = \frac{1}{M} \int \rho_{\rm mass}(r) e^{-i {\bf q} \cdot
{\bf r}} d^3 r = \frac{1}{M} \int_0^\infty \rho_{\rm mass}(r) \,
\frac{\sin qr}{qr} \, 4\pi r^2 dr. \label{eq:fftransform}
\end{equation}

Since the mass distribution in the nucleus is difficult to probe,
it is generally assumed that mass and charge densities are
proportional
\begin{equation}
\rho_{\rm mass}(r) = \frac{M}{Ze} \rho_{\rm charge}(r) ,
\end{equation}
so that charge densities, determined through elastic electron
scattering, can be utilized instead. Because of the normalization
at $q=0$, the proportionality assumption amounts to
\begin{equation}
F_{\rm mass}(q) = F_{\rm charge}(q).
\end{equation}

It is of course convenient to have an analytic expression for the
form factor.  Up to now, this expression has been provided by the
Helm form factor \cite{Helm} given by

\begin{equation}
|F^{SI}(q)|^2 = \left( \frac{3j_1(qR_1)}{qR_1} \right)^2 e^{-q^2
s^2}, \label{eq:HELM}
\end{equation}
where
\begin{equation}
j_1(x) = \frac{\sin x}{x^2} - \frac{\cos x}{x}
\end{equation}
is a spherical Bessel function of the first kind, and where $R_1$
is an effective nuclear radius and $s$ is the nuclear skin
thickness, parameters that need to be fit separately for each
nucleus. The Helm form factor was introduced as a modification of
the form factor for a uniform sphere multiplied by a gaussian to
account for the soft edge of the nucleus (see the next section for
a more complete description).  The parameters $R_1$ and $s$ were
in the past chosen to match numerical integration of Two-Parameter
Fermi (Woods-Saxon) or other parametric models of nuclear density.

In the past, Lewin and Smith \cite{smith} demonstrated a method
for fitting parameters in the Helm form factor to muon
spectroscopy data in the Fricke et al.\ compilation \cite{muon data
new}.  They performed a two-parameter least squares fit to the
muonic spectroscopy data, finding the values of $R_1$ and $s$ for
the Helm form factor which best reproduce the numerical fourier
transform of a Two-Parameter Fermi distribution. Explicitly they set
\begin{equation}
R_1=\sqrt{c^2+\frac{7}{3} \pi^2 a^2 - 5 s^2}
\label{lewinsmith1}
\end{equation}
and take $s\simeq 0.9~{\rm fm}$, $a\simeq 0.52~{\rm fm}$ (as did Fricke et al. in their table IIIA), and
\begin{equation}
 c\simeq 1.23 A^{1/3} - 0.60 ~{\rm fm}
\label{lewinsmith2}
\end{equation}
(which is a least-square fit to the same table in Fricke et al.).
This procedure, however, should be
approached with caution due to the fact that the results depend on
the nuclear density model (in this case Two-Parameter Fermi) which
was used in the original fit to the data.  Also, in the Fricke et
al. compilation \cite{muon data new} in their table IIIA the value of the skin
thickness was fixed, leading to a one parameter fit. Hence the fit
to the form factor generated from the muon spectroscopy data is in
essence a fit to a fit.  The chief advantage here is that a more
accurate Helm form factor is generated which is analytic and
eliminates the need for numerical integration.

It is true that the necessity of introducing a nuclear form factor
correction, particularly in direct dark matter detection, has been
widely recognized and written about in the literature (see for
example \cite{nuclear} and \cite{smith}).  However, what is not
widely known in the dark matter community is that there exist
model independent form factors derived from elastic electron
scattering data that are both analytic (reproducing the chief
advantage of using the Helm form factor) and more accurate than
standard Helm form factors.  These model independent form factors
are derived directly from elastic scattering data, and more
importantly the relevant paramaterizations exist for the large
range of nuclei relevant to dark matter searches (whereas fits to
other paramaterizations such as the harmonic oscillator or
modified harmonic oscillator model exist for only a few nuclei).
We would like to suggest the use of these model independent,
analytic expressions and point out that for many nuclei relevant
to dark matter searches there exist significant differences
between these form factors and the standard Helm parameterization.
A new result which we will present is that these differences in
form factors can lead to 10--20\% shifts in the upper limits
set on the WIMP-proton cross section as published by various
experimental groups.

\section{Nuclear Charge Densities and Form Factors}

Since the interaction between large nuclei and heavy neutralinos
cannot be treated as scattering from point-like particles, it
becomes necessary to introduce models for nuclear charge density
into dark matter detection rate calculations.  In our review of
the literature we found considerable difference in notation
regarding various nuclear charge density/form factor models; we
therefore begin with a brief review of the pertinent
parameterizations.

The simplest example of a model for the charge density of a
nucleus is the uniform model in which the charge density is
constant up to some cut-off radius $R$, i.e.
\begin{equation}
\rho_U(r) = \left\{ \begin{array}{ll} \frac{3Ze}{4\pi R^3},  & r < R, \\
0, & r > R,
\end{array} \right.
\end{equation}
where the charge density is normalized such that the
total charge contained in the nucleus is $Ze$.  The form factor
for this uniform model is simple to calculate and is given by
\begin{equation}
F_U(q) = \frac{3}{qR} j_1(qR).
\end{equation}
Obviously such a
charge density is nonphysical; a nucleus cannot exhibit such an
infinitely sharp cutoff in its charge distribution.

The Helm charge density solves the problem of the infinitely sharp
cutoff in the uniform model by convoluting the uniform charge
density with a Gaussian ``surface smearing" density.  The Helm
charge distribution is given as
\begin{equation}
\rho_H(\vec r) = \int \rho_U(\vec r~') \rho_G(\vec r - \vec r~')
d^3 \vec r~',
\end{equation}
where $\rho_G(\vec r)$ is taken to be
\begin{equation}
\rho_G(\vec r) = \frac{1}{(2 \pi g^2)^{3/2}} e^{-r^2/2g^2}.
\end{equation}
Here $g$ is a parameter which is related to the radius
of the Gaussian smearing surface.  One advantage of the Helm
charge density is that it has an extremely simple analytic form
factor; in fact (by the convolution theorem), the form factor is
simply a product of the form factors of $\rho_U$ and $\rho_G$
\cite{Uberall}:
\begin{eqnarray}
F(q) & = & F_U(q) F_G(q) \nonumber \\
& = & \frac{3}{qR} j_1(qR) e^{-g^2 q^2/2}.
\end{eqnarray}

Returning to dark matter, in most dark matter direct detection
calculations/simulations, the nuclear charge density is assumed
instead to be of the Two-Parameter Fermi (Woods-Saxon)
distribution form given by
\begin{equation}
\rho(r) = \frac{\rho_c}{e^{(r-c)/a} + 1},
\end{equation}
where $c$ is the half-density radius, $\rho_c$ is the density at
$r=c$, and the parameter $a$ is related to the surface thickness
$t$ by $t=(4 \ln 3)a$; the charge density is again normalized by
requiring the total charge contained in the nucleus to be $Ze$.
The Two-Parameter Fermi (Woods-Saxon) distribution has been
favored due to its relative simplicity (only two free parameters)
as well as the fact that parameterizations for many nuclei have
been determined from nuclear scattering experiments. However, the
Fourier transform of the Two-Parameter Fermi or Woods-Saxon
distribution must be calculated numerically since no closed form
analytical Fourier transform exists.  For computational
simplicity, the spin independent form factor is usually taken to
be of the Helm form as described above.  As described in
\cite{nuclear} and \cite{Helm} the Helm form factor is ``virtually
indistinguishable'' from the numerical Woods-Saxon/Two-Parameter
Fermi form factor. For example, DarkSUSY 4.1 \cite{DarkSUSY} takes
the spin independent form factor to be of the Helm form
(\ref{eq:HELM}) with
\begin{equation}
\label{dshelm1}
R_1 = \sqrt{R^2 - 5s^2},
\end{equation}
\begin{equation}
\label{dshelm2}
R=[0.91(M/{\rm GeV})^{1/3} +
0.3]~{\rm  fm},
\end{equation}
and
\begin{equation}
\label{dshelm3}
s=1~{\rm fm}.
\end{equation}
Here the expression for $R$ is that of the nuclear radius in a common parametrization \cite{eder,jkg}, while $s$ is taken from \cite{engel}.

The value of the momentum transfer at which the form factor is
evaluated depends on the recoil energy and the mass of the nucleus
in question.  The momentum transfer can be written as
\begin{equation}
q = \sqrt{2 M E},
\end{equation}
where $M$ is the mass of the target nucleus and $E$ is
the nuclear recoil energy.  The momentum transfer $q$, commonly
given in GeV, can easily be converted to fm$^{-1}$ through
\begin{equation}
q[\textrm{fm}^{-1}] = \frac{\sqrt{2 M[\textrm{GeV}]
E[\textrm{keV}] \times 10^{-6}}}{0.197~\textrm{GeV fm}},
\end{equation}
where the nuclear mass is given in GeV and the nuclear
recoil energy in keV.

\section{Elastic Electron Scattering and Model Independent Form Factors}

Elastic electron scattering as a tool for probing nuclear
structure was pioneered in 1953 at the Stanford Linear Accelerator
by Hofstadter and collaborators \cite{Hofstadter}.  In the next
twenty to thirty years a tremendous amount of data was gathered
which allowed precise determination of nuclear charge distribution
parameters culminating in a 1986 compilation of nuclear charge
density parameters for over one hundred nuclei in Atomic Data and
Nuclear Data Tables \cite{atomic data}.  Many nuclei have charge
densities expressed in multiple parameterizations: two parameter
Fermi, three parameter Fermi, three-parameter gaussian, etc.

Although the Two-Parameter Fermi or Woods-Saxon charge
distribution fits many nuclei well, actual nuclear charge
densities may be more complex than this simple parameterization
allows for.  For example, Figure~1 shows the charge densities for
$^{40}$Ca and $^{208}$Pb plotted versus nuclear radius. $^{40}$Ca
shows an increase in density towards the nuclear center, while
$^{208}$Pb shows a non-constant interior density manifesting in a
depressed charge density at about 3 fm (see for example
\cite{Frauenfelder} and \cite{Sick1974}). These features are not
well-produced by a generic Two-Parameter Fermi or Woods-Saxon
charge distribution. More complex charge densities are therefore
necessary to reproduce these features.

\begin{figure}[t]
\includegraphics[width=0.89\textwidth]{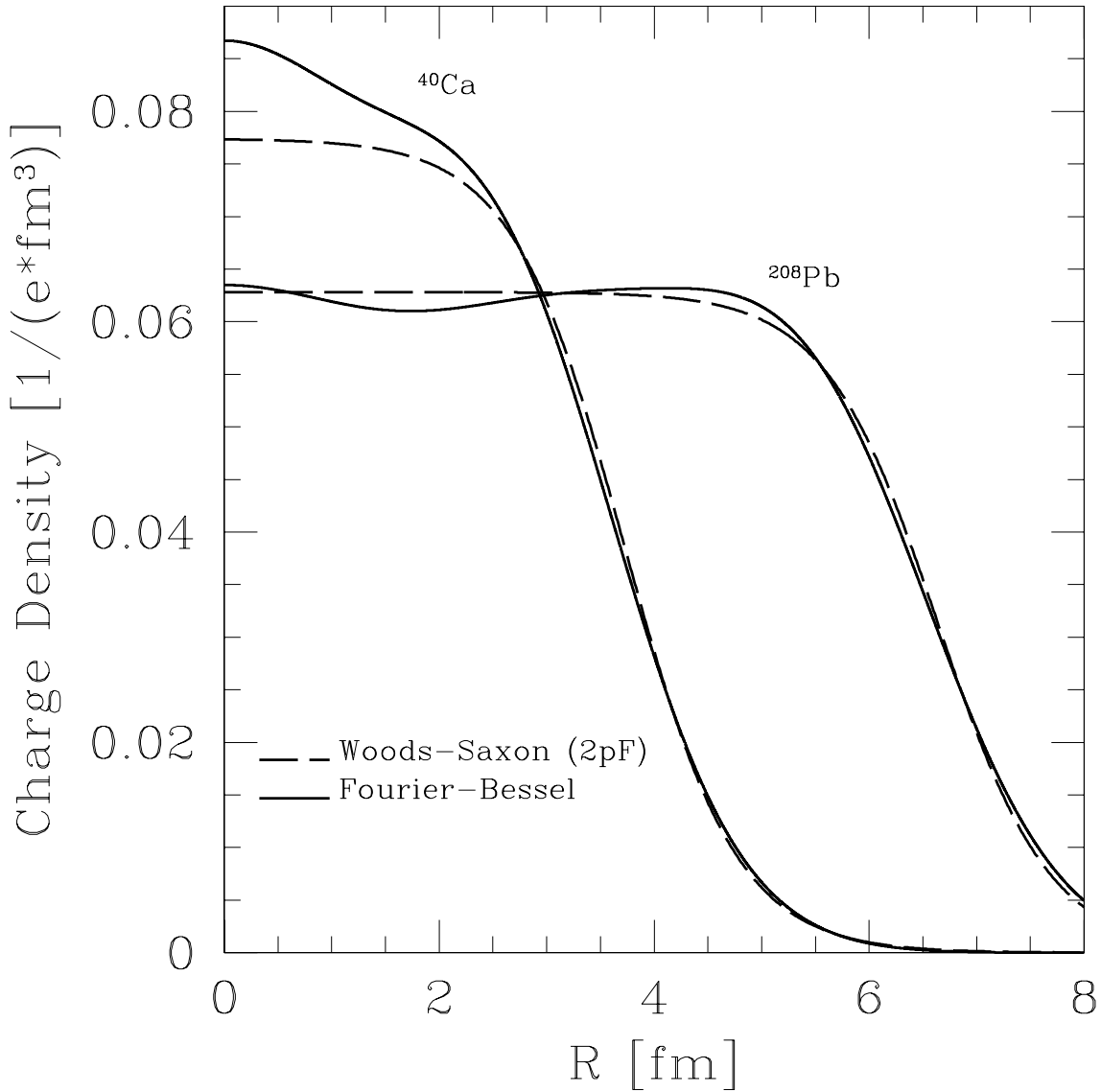}
\caption{Charge density for $^{40}$Ca and $^{208}$Pb as
obtained from elastic electron scattering data.}
\label{fig:1}
\end{figure}

Additionally, most charge densities determined from electron
scattering experiments have been analyzed using model
distributions such as the Two-Parameter Fermi or Woods-Saxon
distributions.  The drawback here is that charge density
parameters measured are fundamentally model dependent.  This point
applies as well to harmonic oscillator or modified harmonic
oscillator charge density parameterizations, despite the fact that
these have been used to characterize a limited number of nuclei
involved in dark matter searches.   As noted by Sick
\cite{Sick1974}, it is at times unclear whether the charge density
parameters measured are the fundamental characterization of the
nuclear charge density or if they are simply the result of fitting
data to a model which may be too restrictive in form.  To combat
this fundamental uncertainty, model independent methods for
analyzing electron scattering data and extracting nuclear charge
densities were developed by several groups.  Of course, these
methods were not completely model independent; some model
dependence is necessary since electron scattering data is only
available over a finite momentum transfer range.  Two primary
methods emerged for extracting nuclear charge density parameters
in a model independent fashion: an approach in which the charge
density is written as a sum of gaussians and an alternate approach
in which the charge density is written as a Fourier superposition
of Bessel functions.

In the Sum of Gaussians expansion (SOG), first introduced by Sick
\cite{Sick1974}, the charge density of a nucleus is modelled as a
series of Gaussians.  The width of the Gaussians, given by the
parameter $\gamma$, is calculated by equating it to the smallest
width of the peaks in the nuclear wave functions as found in
Hartree-Fock calculations; negative amplitudes for the Gaussians
are not allowed as to avoid creating structures with widths
smaller than $\gamma$ due to interference.   As noted in
\cite{atomic data} this method has several advantages: values of
$\rho(r)$ at different radii are decoupled due to the rapid
fall-off of the Gaussian tail, and as long as a sufficient number
of Gaussians are used to ensure a good fit, the results of the
analysis is independent of the number of Gaussians.

In the SOG expansion the charge density is written as
\begin{equation}
\rho(r) = \sum_{i=1}^N A_i \left\{ e^{-[(r-R_i)/\gamma]^2} +
e^{-[(r+R_i)/\gamma]^2} \right\},
\end{equation}
where the coefficients $A_i$ are given by
\begin{equation} A_i = \frac{Z e Q_i}{2 \pi^{3/2} \gamma^3 \left(1+2
R^2_i/\gamma^2 \right)}.
\end{equation}
The $Q_i$ represent the fractional charge carried in the
$i^{\rm th}$ gaussian and lead to the following definition for
the normalization of the charge density:
\begin{equation}
\sum_{i=1}^N Q_i = 1.
\end{equation}

An analytical form factor can be determined for this density
parameterization, which eliminates the necessity of performing
numerical integration to find the form factor as in the
Two-Parameter Fermi density parameterization.  Assuming spherical
symmetry, the form factor in the SOG expansion is given by
\begin{equation}
\label{eq:SOG}
F(q) = e^{-q^2 \gamma^2/4} \sum^N_{i=1} \frac{Q_i}{1+2 R^2_i /
\gamma^2} \left[ \cos(q R_i) + \frac{2 R^2_i}{\gamma^2}
\frac{\sin(q R_i)}{q R_i} \right].
\end{equation}

The other model independent parameterization of nuclear charge
densities that was developed is a Fourier-Bessel expansion. In the
Fourier-Bessel expansion (FB), first introduced by Dreher et al.
\cite{dreher}, the charge density is modelled as a sum of Bessel
functions up to some cut-off radius $R$, and is assumed to be zero
thereafter.  The form factor is assumed to fall off for large $q$
as $q^{-4}$ and $e^{-aq^2}$; these result from the distribution of
nucleons in the nucleus and from the finite extension of the
nucleons respectively.  Although the results depend slightly on
the value of $R$, the cut-off radius, an advantage of this
approach is that this collection of assumptions gives an upper
limit on the contribution of higher Fourier components to the
series expansion \cite{atomic data old}.

Specifically,
\begin{equation}
\rho(r) = \left\{ \begin{array}{ll} \sum_{\nu=1}^N a_\nu j_0 \left(\nu
\pi r / R \right) & \textrm{for}~r \le R, \\ 0 & \textrm{for}~r \ge
R, \end{array} \right.
\end{equation}
where $j_0(x)=\sin x/x$ is the zero-th order spherical Bessel
function.  Here the charge density is normalized by requiring
\begin{equation}
\int \rho(r) d^3 r = Z e.
\end{equation}
This charge density also has the advantage that an
analytical expression may be found for the form factor. Assuming
spherical symmetry, Fourier transforming the charge density yields
\begin{equation}
\label{eq:FB}
F(q) = \frac{\sin(qR)}{qR} \, \frac{ \displaystyle \sum_{\nu=1}^N \frac{ \left(-1\right)^{\nu} a_\nu}{\nu^2\pi^2 - q^2 R^2} }{ \displaystyle \sum_{\nu=1}^N \frac{ \left(-1\right)^{\nu} a_\nu}{\nu^2 \pi^2} },
\end{equation}
which is normalized to $F(0) = 1$.

In a final point regarding model independent charge densities, it
is difficult to estimate the theoretical errors involved in fits
to elastic electron scattering data. As noted in \cite{atomic
data}, in the case of the FB approach the errors in the individual
coefficients are not presented in data compilations since errors
are strongly correlated and can only be extracted from the full
correlation matrix; as this matrix is never published in papers,
it is impossible to give errors on individual coefficients of the
FB expansion for the charge density. Recently, however, Anni, Co',
and Pellegrino in \cite{anni} analyzed model independent
extraction of charge density parameters from elastic electron
scattering data to determine uncertainties and the minimum number
of expansion coefficients needed to give an accurate
representation of the data.  Anni et al. determined that in the
model independent approach a truncation error is unavoidable;
however, they developed an approach to determine an optimal number
of coefficients.  They point out that distributions extracted from
these models must be used with care; however, these models are
still much more robust than model dependent fits such as the
Woods-Saxon or Two-parameter Fermi model.

\section{Nuclear Charge Density Parameters from Muonic Atom
Spectroscopy}

One can also extract nuclear charge parameter factors from nuclei
using muonic atom spectroscopy, however parameters extracted in
this manner are to some degree model dependent as the final
parameter fitting depends on the choice of an analytical charge
density. The nuclear charge distributions are extracted by
considering finite size effects to the energy shift in first-order
perturbation theory (see \cite{muon data old} and \cite{muon data
new}, for example, for a review).  The analysis takes advantage of
Barret moments \cite{barret moments} which can be extracted in a
model independent fashion from the transition energies.  Charge
density parameters are determined by finding the eigenvalues of
the Dirac equation with the analytic charge density fitted to the
experimental transition densities \cite{muon data new}.  In the
analysis of muonic atom spectroscopy data the Two-Parameter Fermi
charge distribution is used:
\begin{equation}
\rho(r) = \frac{\rho_c}{e^{(r-c)/a} + 1}. \label{eq:2ptFermi}
\end{equation}
The skin thickness $t = 4a \ln(3)$ is fixed to 2.30 fm, and the
half-density radius parameter $c$ is fitted to reproduce the
experimental transition energies. This method can also be applied
to deformed nuclei by writing the half-density radius $c$ as
\begin{equation}
c=R_0 \left[1 + \beta_2 Y_{20}(\theta,\phi)\right],
\end{equation}
where $\beta_2$ is the quadrupole deformation parameter,
and $R_0$ is the monopole radius.

As mentioned previously, the disadvantage to the Two-Parameter
Fermi charge density lies in the fact that it has no analytic
Fourier transform, and thus form factors must be computed through
numerical integration.  However, as noted in \cite{smith}, data
from muonic spectroscopy parameterizing the nuclear charge density
in terms of a Two-Parameter Fermi distribution exists for several
nuclei relevant to direct dark matter detection experiments such
as Na, Xe, and I.  Two-Parameter Fermi charge distribution
parameters also exist for $^{184,186}$W. However, these parameters
were obtained from elastic electron scattering experiments. In
order to determine the impact of using the Helm versus the form
factor associated with the Two-Parameter Fermi charge distribution
(and to compare against utilization of the Sum of Gaussian or
Fourier-Bessel parameterizations) we numerically fourier transform
the Two-Parameter Fermi distribution obtaining the associated form
factor, generically referring to it as a Woods-Saxon form factor.
Numerical integration was performed with a 48-point Legendre-Gauss
quadrature.  Although integrating the oscillating $\sin(qr)$ in
~\ref{eq:fftransform} is notoriously subtle, the 48-point
Legendre-Gauss qaudrature used gave robust results and reproduced
form factors previously published in the literature (for example,
see \cite{smith}).  The half-density parameter $c$, the parameter
$a$ (which is related to the nuclear skin thickness $t$), and the
rms value of the charge density for the nuclei of interest are
included the the appendix as Table 7.

\section{Discussion and Results}

To determine the effect of using model independent form factors on
dark matter direct detection rates, DarkSUSY was modified to
compute Helm form factors using the standard parameterizations
from \cite{eder}-\cite{engel} (referred to here onward as Helm/DarkSUSY-4.1) and the Helm form
factors of Lewin and Smith \cite{smith} (referred to here onward
as Helm/Lewin-Smith), as well as model
independent form factors derived from the SOG or FB approach. The
form factors were evaluated at the appropriate momentum transfers
for nuclear recoils between 10 and 100 keV. The 1986 compilation
of electron scattering data \cite{atomic data} was used to
calculate the relevant charge densities and form factors. Nuclei
were chosen based on their importance to current and
future/planned dark matter direct detection searches as well as
the availability of electron scattering data in the appropriate
momentum transfer range.  We have restricted our attention to spin
independent scattering of neutralinos from nuclei.  This is not
too restrictive since many of the nuclei relevant in direct dark
matter searches are even-even nuclei for whom the spin dependent
cross section vanishes in the $J=0$ ground state (see for example
\cite{nuclear}). Since the relevant nuclei tend to be spherical or
roughly spherical in shape, their charge densities and form
factors are able to be analyzed using the SOG or FB approach.
Iodine and Sodium (important to the DAMA experiment; see for
example \cite{DAMA}), Xenon (important to the ZEPLIN and XENON
experiments; see for example \cite{ZEPLIN} and \cite{XENON}), and
Tungsten (important in the CRESST experiment; see for example
\cite{CRESST}) are analyzed using the Two-Parameter Fermi charge
density as in \cite{smith} for the sake of completeness.

A range of nuclei from $^{12}$C to $^{208}$Pb were analyzed to
give a general trend for increasing nuclear mass.  Of particular
importance to dark matter experiments are the following nuclei:
$^{28}$Si (used in the CDMS experiments \cite{CDMS}), $^{32}$S
(used in the DRIFT experiment \cite{DRIFT}), $^{40}$Ca (used in
the CRESST experiments \cite{CRESST}), and $^{70-74}$Ge (used in
the CDMS, EDELWEISS, GENIUS, and CryoArray experiments
\cite{CDMS}-\cite{CryoArray}). For a convenient summary chart of
current and future-planned dark matter direct detection
experiments and the associated target materials see Tables 1-3 in
\cite{Gaitskell}.

\begin{figure}[t]
\includegraphics[width=0.89\textwidth]{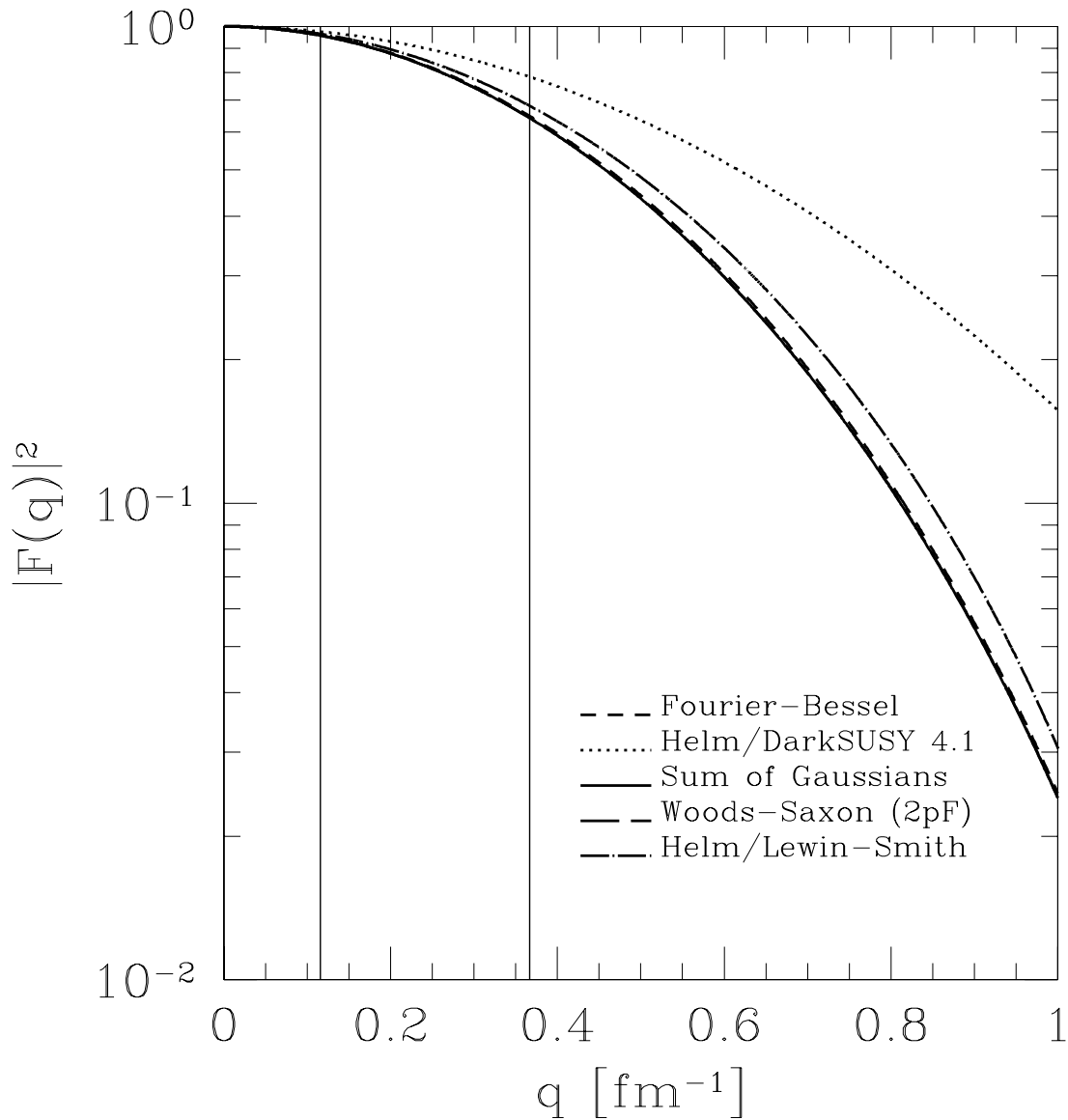}
\caption{Helm, FB, and SOG form factors for $^{28}$Si versus
nuclear radius as obtained from elastic electron scattering data.
Lines for momentum transfers corresponding to 10 and 100 keV
nuclear recoil are included.}
\label{fig:2}
\end{figure}

In Figure \ref{fig:2} we plot $|F(q)|^2$ for $^{28}$Si, an
important target medium in the CDMS experiments, for small
momentum transfers less than 1 fm$^{-1}$.  Lines indicate the
momentum transfers which corresponds to 10 keV and 100 keV nuclear
recoils. As can be seen from the plot, all of the form factors in
their various parameterizations are not significantly different at
low ($\sim$10 keV) nuclear recoils (they differ by at most 2\%),
and the Helm form factor (in either parameterization) may be used
with confidence. However, for larger momentum transfers the FB,
SOG, and Woods-Saxon form factors begin to diverge from the Helm form factors; at a momentum transfer
corresponding to 100 keV there is a 20.9\% difference between the
Helm/DarkSUSY-4.1 and the FB/SOG form factors. The Helm form factor of
Lewin and Smith differs from the FB or SOG form factors negligibly
at small momentum transfers and up to 5\% at 100 keV nuclear recoil.
Hence for small nuclei at low momentum transfers all
parameterizations of the form factor fit well.  One should also
note, however,  that the FB, SOG, and WS form factors, despite the
different parameterizations, are virtually indistinguishable,
giving further confidence in the use of model independent form
factors.  The Woods-Saxon form factor is a numerical fourier
transform of a physical charge density model (the Two-Parameter
Fermi), and should be in some sense thought of as the closest approximation
to the actual form factor. The two model independent form factors (FB and SOG)
trace the nuclear density more accurately, and easier to use as they are analytic: they should be preferred whenever available.

\begin{figure}[t]
\includegraphics[width=0.89\textwidth]{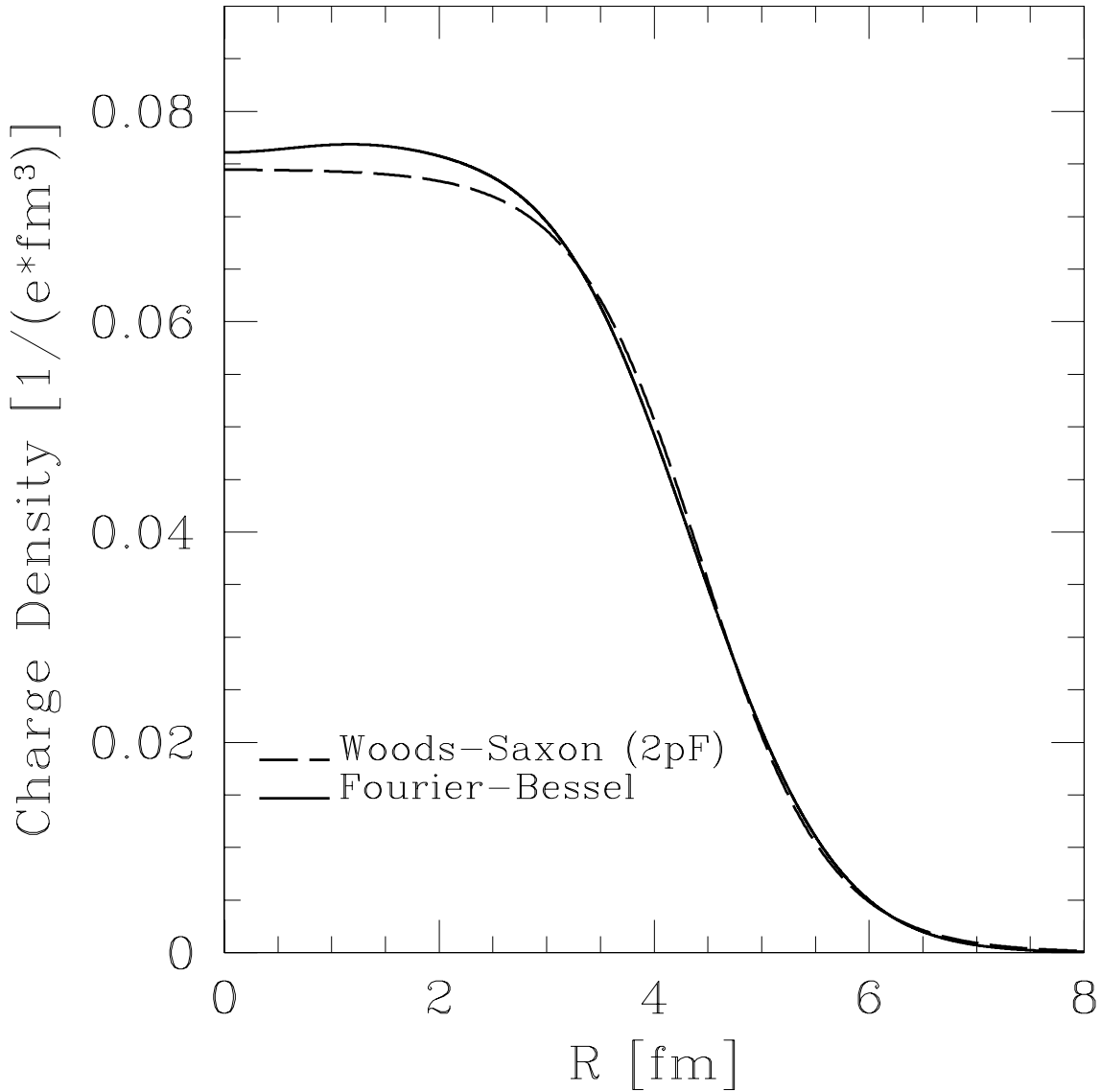}
\caption{Charge density for $^{70}$Ge in the Woods-Saxon and
FB parameterizations.}
\label{fig:3}
\end{figure}

\begin{figure}[t]
\includegraphics[width=0.89\textwidth]{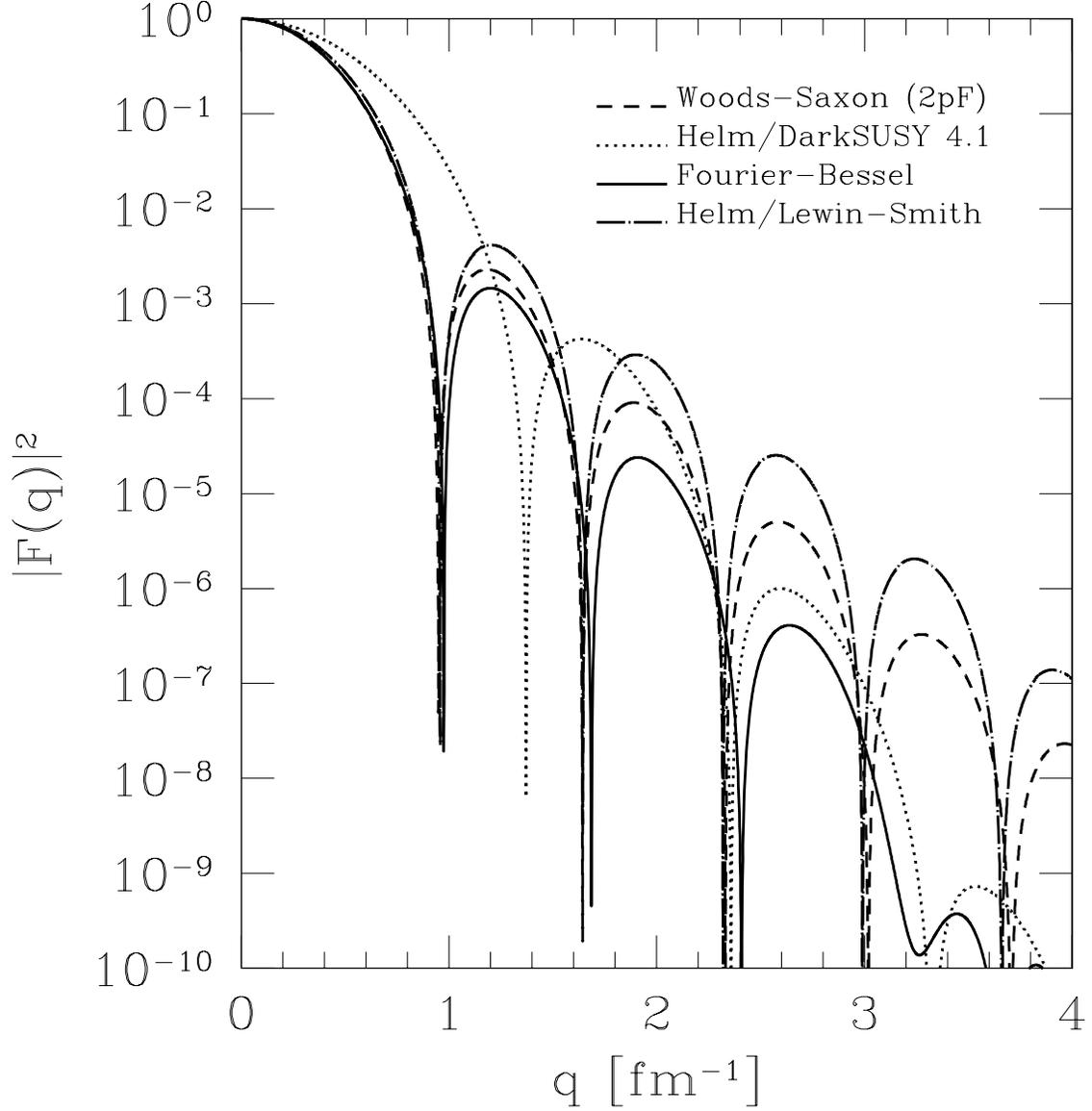}
\caption{Helm and FB form factors for $^{70}$Ge.}
\label{fig:4}
\end{figure}

\begin{figure}[t]
\includegraphics[width=0.89\textwidth]{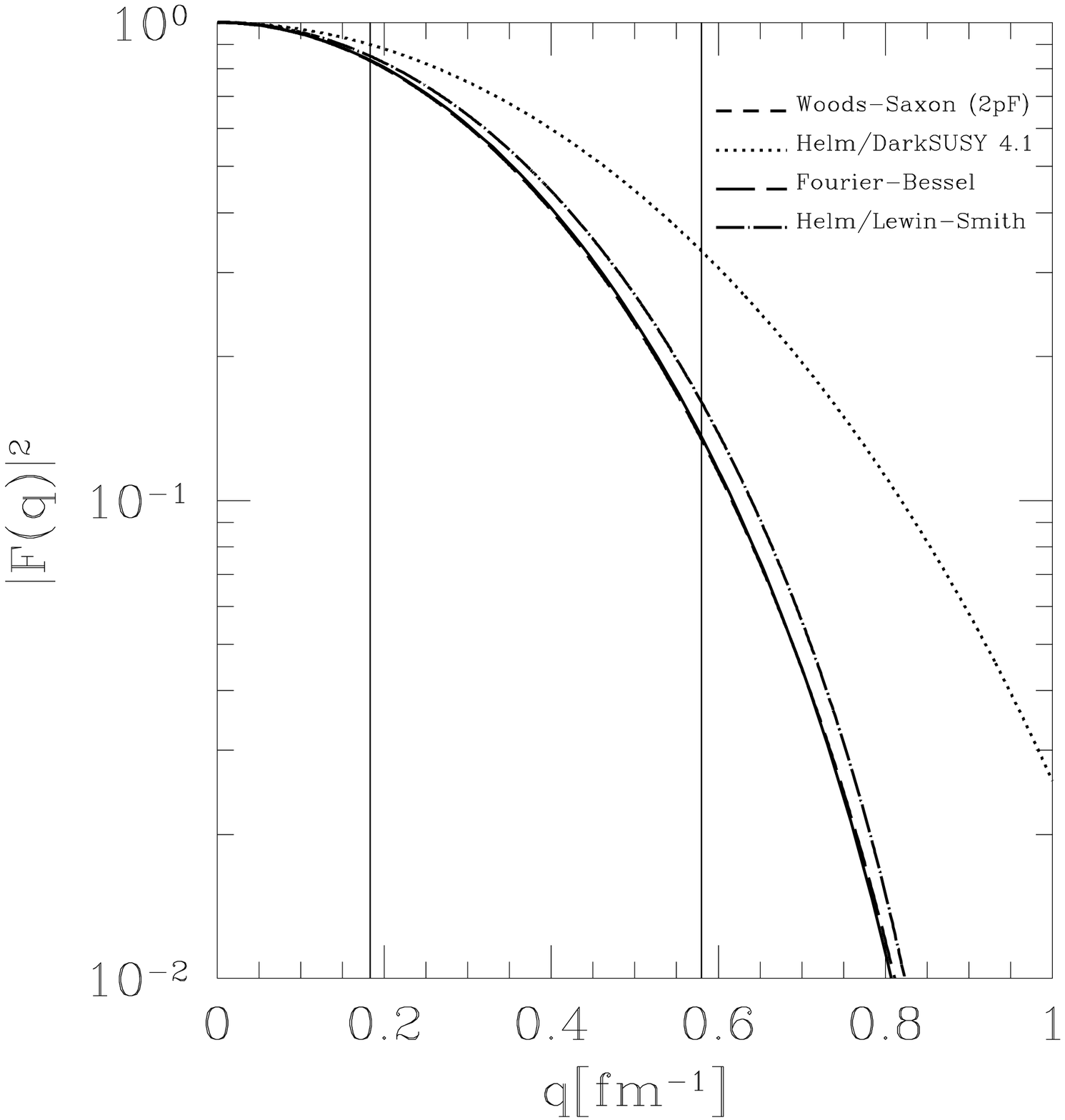}
\caption{Helm and FB form factors for $^{70}$Ge with
10 and 100 keV recoil lines for small momentum transfers.}
\label{fig:5}
\end{figure}

In Figure \ref{fig:3} we plot the charge density for $^{70}$Ge
using both model independent fits to elastic electron scattering
data (the FB fit) and also using the standard Woods-Saxon
(Two-Parameter Fermi) charge density.  As in Figure \ref{fig:1},
the charge density for germanium shows a non-constant interior
density which cannot be characterized entirely by the Woods-Saxon
distribution. Figure \ref{fig:4} shows the Helm/DarkSUSY-4.1,
Helm/Lewin-Smith, Woods-Saxon, and FB form factors for $^{70}$Ge
plotted over a range of 0 to 4 fm$^{-1}$. As can be seen from the
plot, the Helm/DarkSUSY-4.1 and the other form factors differ
significantly over the range of momentum transfers. Most
significantly, the first diffraction minimum in the FB and WS form
factors occurs close to 1 fm$^{-1}$ whereas the first minimum in
the Helm/DarkSUSY-4.1 form factor occurs at significantly larger
momentum transfer. The difference in form factors can be traced to
the use of the nuclear radius formula (\ref{dshelm2}), from
\cite{eder}, instead of the more accurate one in
Eq.~(\ref{lewinsmith1}), from \cite{smith}.

\begin{figure}[t]
\includegraphics[width=0.89\textwidth]{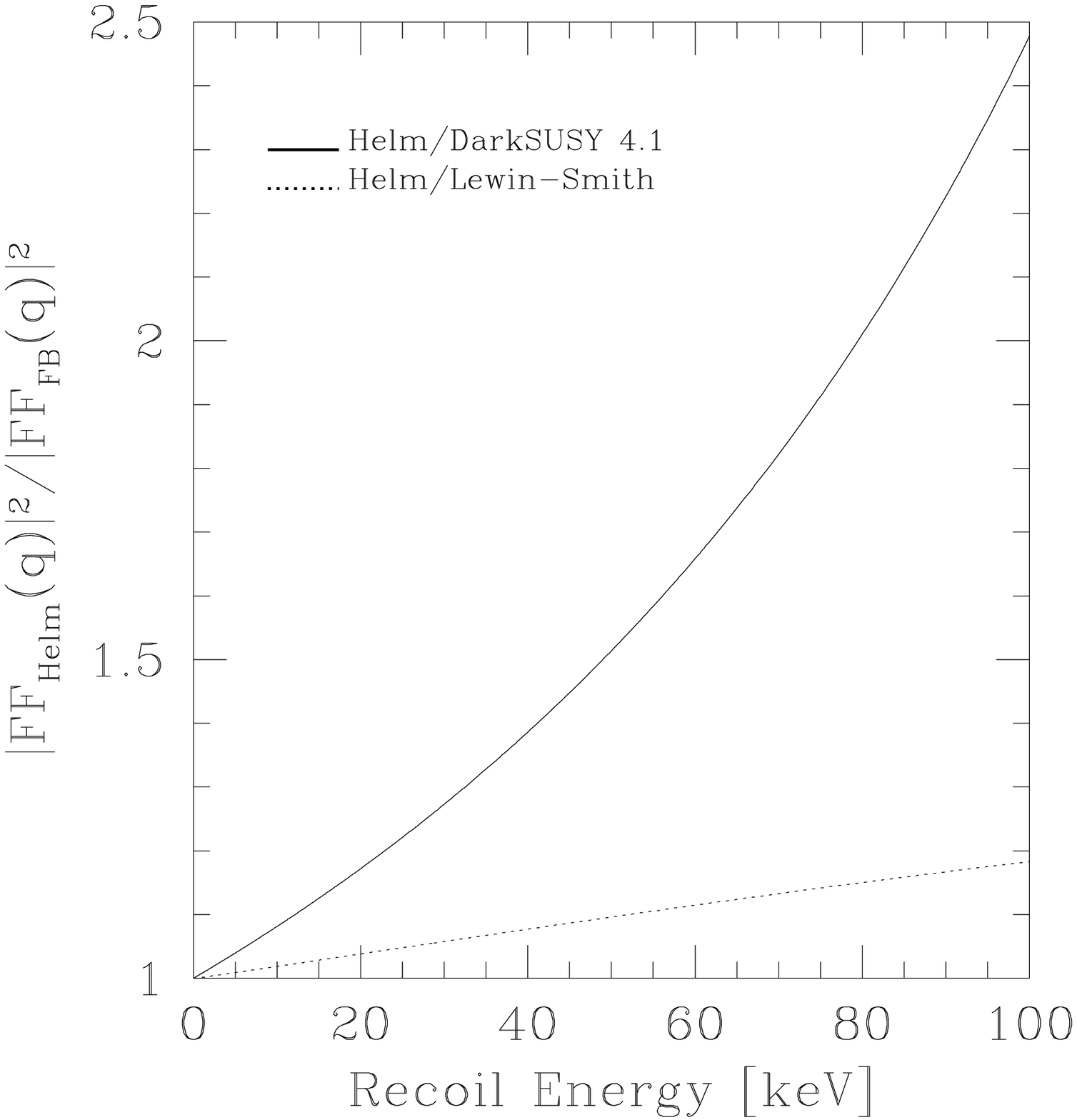}
\caption{Ratio of the form factors squared in the helm and
Fourier Bessel schemes for $^{70}$Ge.}
\label{fig:6}
\end{figure}

 Figure \ref{fig:5}
shows the same plot of form factors for $^{70}$Ge but this time
highlighting the relevant range of momentum transfers important
for direct detection experiments. The momentum transfers
corresponding to 10 and 100 keV nuclear recoils are highlighted by
vertical lines. Although the divergence between the two form
factors at 10 keV nuclear recoil is small (8.12\% exactly between
Helm/DarkSUSY-4.1 and FB form factors and 2.0\% for the
Helm/Lewin-Smith and the FB form factors), the form factors are
already beginning to diverge. For larger nuclear recoils the form
factors continue to diverge leading to a 27.4\% and 6.3\%
difference at 30 keV, a 65.8\% and 12.1\% difference at 60 keV,
and a 148.0\% and 19.6\% difference at 100 keV for the
Helm/DarkSUSY-4.1 vs. the FB  and the Helm/Lewin-Smith vs. the FB
form factor respectively. Again the WS and FB form factors are
indistinguishable over the relevant momentum transfer range. For
illustrative purposes, we plot the ratio of the form factor
squared in the Helm scheme to that of the form factor squared in
the FB expansion for $^{70}$Ge in Figure \ref{fig:6}.

\begin{figure}[t]
\includegraphics[width=0.89\textwidth]{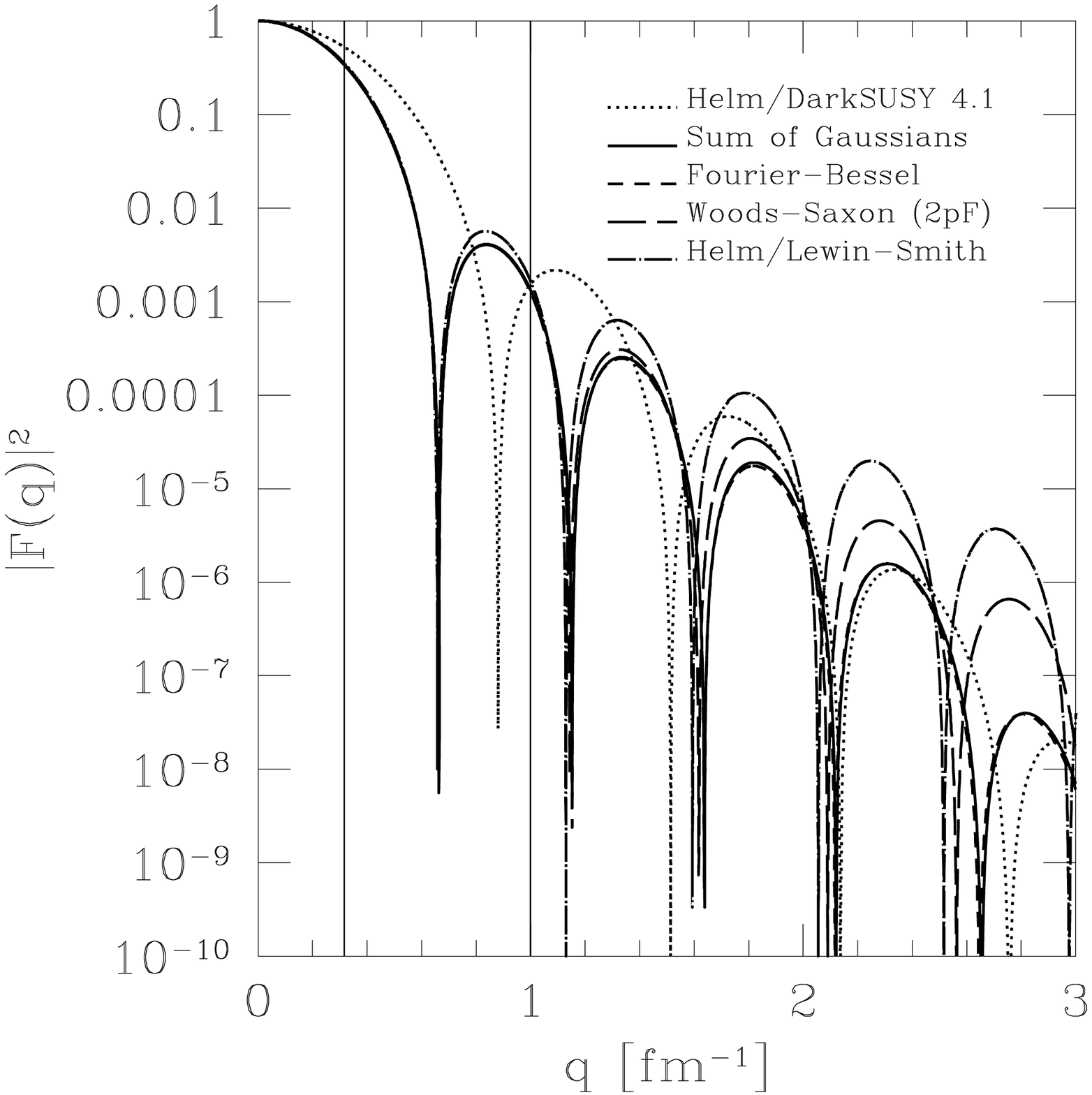}
\caption{Helm and FB form factors for $^{208}$Pb with
10 and 100 keV recoil lines for momentum transfers less than 3.0
fm$^{-1}$.}
\label{fig:7}
\end{figure}

In Figure \ref{fig:7} we plot $|F(q)|^2$ for $^{208}$Pb for
momentum transfers in the range of zero to about 2 fm$^{-1}$.
Although no direct dark matter searches currently employ lead as a
detection medium (though it is used for shielding), lead is an
excellent example of the general trend for heavier nuclei.  As the
atomic number and mass of the nucleus increases, the Helm/DarkSUSY-4.1, FB, SOG,
and WS form factors begin to diverge at smaller
momentum transfers.  As can be seen in the case of $^{208}$Pb, the
first diffraction minimum for the FB form factor occurs between
0.6 and 0.7 fm$^{-1}$ whereas the first diffraction minimum for
the Helm/DarkSUSY-4.1 form factor occurs for the much larger momentum
transfer of about 0.9 fm$^{-1}$.  The Helm form factor of
Lewin and Smith does a much better job matching the WS, FB, and
SOG form factors, however the percent difference can become quite
large (as high as 41\% as 60 keV recoil energy).  This discrepancy
leads to a large difference in the spin independent cross sections
of neutralinos with $^{208}$Pb for nuclear recoils as small as 10
keV. Although these diffraction minima in electron or muon
scattering can be partially filled by multiple photon exchanges in
the nucleus, in the case of neutralinos this is not expected to
occur \cite{smith}. Lead clearly illustrates the hazards of using
large-A nuclei as a detection medium; potential signals may be
cut-off due to very small form factors in the relevant range of
momentum transfers.

In summary, Table 1 gives the ratios the Helm form factor squared
to either FB, SOG, or WS form factors squared for various nuclei
important in dark matter searches at 10, 30, 60, and 100 keV
nuclear recoils. Note that $^{27}$Al is unique in that its Helm form factor as computed in DarkSUSY 4.1 is actually
smaller than the corresponding model independent form factor;
expected detection rates for an Al target should therefore be
higher than previously expected.  Table 2 gives the ratio of the
Helm/Lewin-Smith form factor squared to either the FB, SOG, or
WS form factors squared for various nuclei at a series of nuclear
recoil energies.

\begin{table}[t]
\begin{center}
\begin{tabular}{|c|c|c|c|c|}
\hline Nucleus & $F^2(q)_{\rm H}/F^2(q)$  &
$F^2(q)_{\rm H}/F^2(q)$  &
$F^2(q)_{\rm H}/F^2(q)$  & $F^2(q)_{\rm H}/F^2(q)$ \\
& 10 keV & 30 keV & 60 keV & 100 keV\\
\hline $^{12}$C & 1.01 & 1.01 & 1.03 & 1.06 \\
\hline $^{16}$O & 1.01 & 1.03 & 1.06 & 1.10  \\
\hline $^{23}$Na & 1.02 & 1.05 & 1.10 & 1.17 \\
\hline $^{27}$Al & 0.99 & 0.97 & 0.94 & 0.90 \\
\hline $^{28}$Si & 1.02 & 1.06 & 1.12 & 1.21 \\
\hline $^{32}$S & 1.02 & 1.08 & 1.16 & 1.28 \\
\hline $^{40}$Ar & 1.03 & 1.10 & 1.22 & 1.41 \\
\hline $^{40}$Ca & 1.03 & 1.11 & 1.23 & 1.42  \\
\hline $^{70}$Ge & 1.08 & 1.27 & 1.66 & 2.48  \\
\hline $^{127}$I & 1.21 & 1.89 & 4.98 & 13739.7$^\star$\\
\hline $^{129}$Xe & 1.22 & 1.92 & 5.36 & 1164.1$^\star$ \\
\hline $^{134}$Xe & 1.22 & 1.96 & 5.91 & 100.8$^\star$ \\
\hline $^{184}$W & 1.48 & 4.68 & 20.80  & 0.02 \\
\hline $^{208}$Pb & 1.54 & 6.91 & 1.64 & 1.15 \\
\hline
\end{tabular}
\end{center}
\caption{Ratios of the Helm/DarkSUSY 4.1 \cite{DarkSUSY} to FB or
SOG or WS form factors squared for various nuclei important in
current or future direct dark matter searches at 10, 30, 60 and
100 keV nuclear recoil. Lead, though not important in direct dark
matter searches, is added as an example of a very large $A$
nucleus.} \label{tab:1}
\end{table}

\begin{table}[t]
\begin{center}
\begin{tabular}{|c|c|c|c|c|}
\hline Nucleus & $F^2(q)_{\rm H}/F^2(q)$  & $F^2(q)_{\rm
H}/F^2(q)$  &
$F^2(q)_{\rm H}/F^2(q)$  & $F^2(q)_{\rm H}/F^2(q)$ \\
& 10 keV & 30 keV & 60 keV & 100 keV\\
\hline $^{12}$C & 1.00 & 1.00 & 1.01 & 1.01 \\
\hline $^{16}$O & 1.00 & 1.01 & 1.02 & 1.03  \\
\hline $^{23}$Na & 1.01 & 1.02 & 1.03 & 1.05 \\
\hline $^{27}$Al & 1.00 & 1.01 & 1.02 & 1.04 \\
\hline $^{28}$Si & 1.00 & 1.01 & 1.03 & 1.05 \\
\hline $^{32}$S & 1.01 & 1.02 & 1.05 & 1.08 \\
\hline $^{40}$Ar & 1.01 & 1.03 & 1.05 & 1.09 \\
\hline $^{40}$Ca & 1.01 & 1.03 & 1.06 & 1.1  \\
\hline $^{70}$Ge & 1.02 & 1.06 & 1.12 & 1.19  \\
\hline $^{127}$I & 1.03 & 1.08 & 1.13 & 1.27 \\
\hline $^{129}$Xe & 1.03 & 1.08 & 1.13 & 2.98 \\
\hline $^{134}$Xe & 1.02 & 1.05 & 1.03 & 2.67 \\
\hline $^{184}$W & 1.07 & 1.25 & 0.99 & 1.6 \\
\hline $^{208}$Pb & 1.03 & 1.01 & 1.41 & 1.23 \\
\hline
\end{tabular}
\end{center}
\caption{Ratios of the Helm/Lewin-Smith form factor \cite{smith}
squared to FB or SOG or WS form factors squared for various nuclei
important in current or future direct dark matter searches at 10,
30, 60 and 100 keV nuclear recoil. Lead, though not important in
direct dark matter searches, is added as an example of a very
large $A$ nucleus.} \label{tab:2}
\end{table}

In Table~\ref{tab:1}, the starred entries show an extreme
difference between the Helm and other form factors in which the
ratio is greater than one hundred; this occurs when the form
factor is evaluated on or about a diffraction minimum for the WS,
FB, or SOG form factors with the corresponding Helm/DarkSUSY-4.1
form factor diffraction minimum occurring at higher momentum
transfer. Note that no such entries occur in Table 2 as the
Helm/Lewin-Smith form factor has diffraction minima whose
placement roughly matches the locations in the FB, SOG, or WS form
factors. As can be seen from Tables \ref{tab:1} and \ref{tab:2},
although either Helm form factor may be used with confidence for
light nuclei ($A < 30$) at low nuclear recoils, model independent
form factors become increasingly more accurate for large nuclei
even at relatively modest momentum transfers. Failure to account
for this correction to the form factor can lead to 10--20\% errors
in cross sections and detection rates.

\begin{figure}[t]
\includegraphics[width=0.89\textwidth]{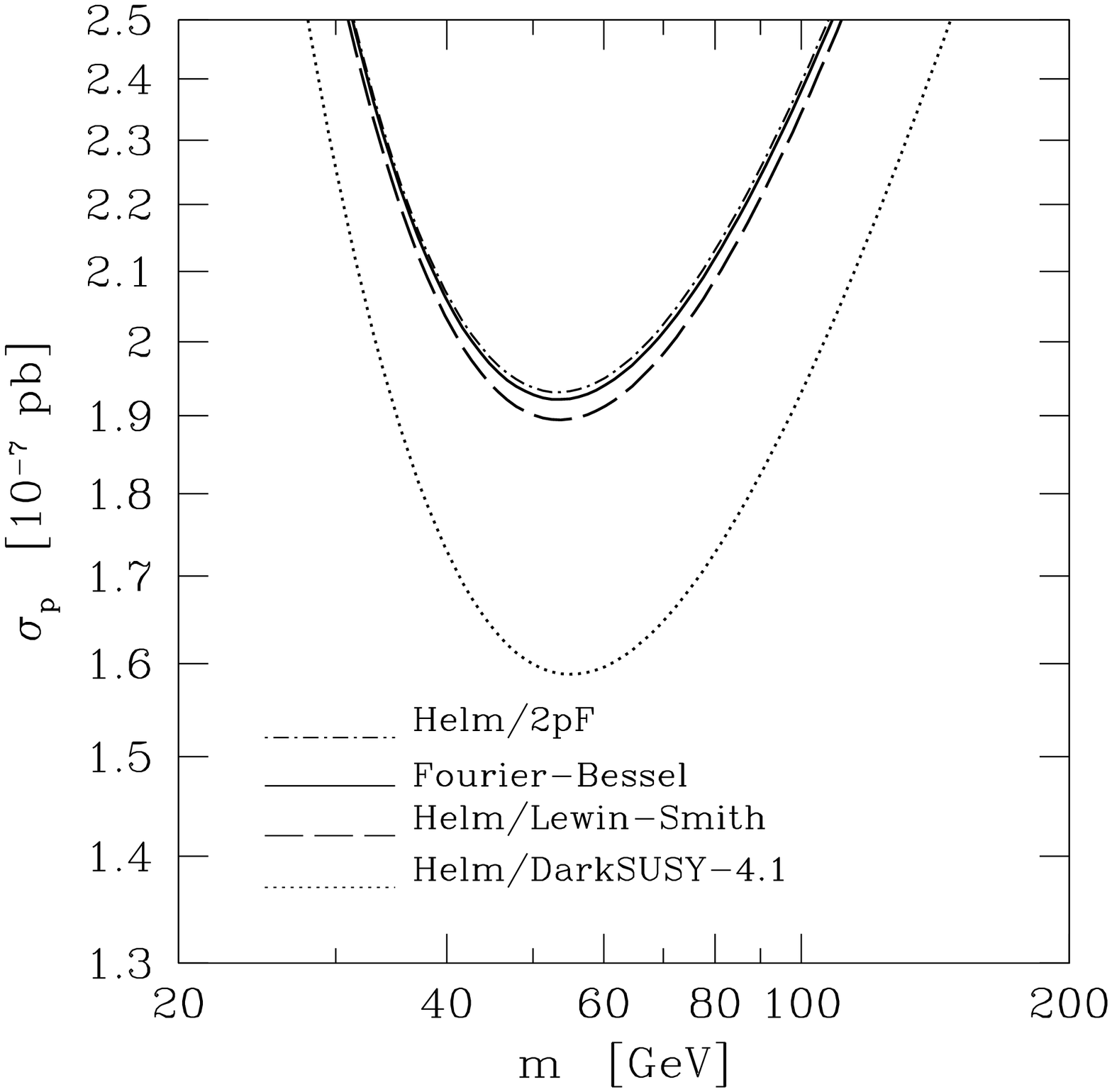}
\caption{Hypothetical upper limit on the WIMP-proton cross section for a
100 kg-day exposure of $^{70}$Ge (assuming an efficiency of 50\% and no detected event)
and various form factors.} \label{fig:8}
\end{figure}

We want to illustrate the importance of using accurate form
factors when setting limits on WIMP-proton cross sections. For
this purpose, we imagine a hypothetical detector made of
$^{70}$Ge. We assume a simple isothermal sphere model for the
WIMPs in the Milky Way halo, so that the WIMP speeds with respect
to the halo obey a Maxwellian distribution with a velocity
dispersion truncated at the escape velocity. We use the same halo
velocity distribution function as \cite{sagittarius stream}  --
see particularly (30)-(33) -- setting $\sigma_h=220$ km/s,
$v_E=220$ km/s, $v_{\rm esc}=650$ km/s, and $\rho=0.3$
GeV/$c^2$/cm$^3$.  We assume a detector energy threshold of 10
keV, an efficiency of 50\%, an exposure of 100 kg-days, and no
events observed in the energy range of 10-50 keV.

In Figure \ref{fig:8} we show the upper limit on the WIMP-proton
cross section obtained using (\ref{sigmasi}-\ref{eq:2}) with
$N_{10-50 {\rm keV}} < 2.3$ and various models of the form factor.
The form factor as given by DarkSUSY 4.1 is a Helm type
form factor calculated with the parameters given in
(\ref{dshelm1}-\ref{dshelm3}). The form factor computed using the procedure in
Lewin and Smith \cite{smith} is a Helm type form factor
with parameters given by (\ref{lewinsmith1}-\ref{lewinsmith2}).
The Fourier Bessel form factor is calculated as in Equation
\ref{eq:FB} using the data given in Appendix I Table 4. Finally,
the form factor labelled ``Helm/2pF'' is a Helm type form factor computed using
$R_1$ in (\ref{lewinsmith1}) with $s = 0.9$ fm and with $a=0.5807$ fm and $c=4.430$
fm directly from the Two-Parameter Fermi distribution in table VIII in Fricke et al.\
(instead of the least-square fit to their table IIIA in Lewin and Smith).

In this example the limits computed with DarkSUSY 4.1 differ from
the Fourier Bessel limits by 10--20\%, while those computed a la
Lewin and Smith differ from the Fourier Bessel limits by 1--2\%.
(Notice that the logarithmic vertical scale covers a range much
smaller than the usual one in this kind of plots.) Numerically, we
can compare our hypothetical upper limit on the WIMP-proton cross
section for a 100 GeV WIMP, say. Using the Fourier Bessel form
factor we find the limit to be $2.38 \times 10^{-7}$ pb.
Using the Helm form factor with $a$ and $c$ parameters directly from Fricke et al.\
we find it to be $2.39 \times 10^{-7}$ pb, which is 0.5\% larger.
Using the Helm form factor a la Lewin and Smith we find the limit to be
$2.34 \times 10^{-7}$ pb, which is 1.5\% smaller. Using the Helm
form factor computed with DarkSUSY 4.1 we find the upper limit to
be $1.93 \times 10^{-7}$ pb, which is 19\% smaller.

The improvement in form factors that incorporate nuclear data more
and more accurately is quite evident in this example.  The upper
limits on the WIMP-proton cross section calculated using form
factors generated with the Helm distribution utilizing muonic
spectroscopy data compiled in Fricke et al. \cite{muon data new}
differ only by less than 1\% from the WIMP-proton cross section
calculated using the Fourier Bessel form factor. The Helm form
factor computed a la Lewin and Smith is also a very good
approximation to the Fourier Bessel form factor, to within a few
percent. The Helm form factor in DarkSUSY 4.1 is further off, by
10--20\%, because of a poor choice of the parameter $R_1$, which
is there set equal to the nuclear radius in standard
parameterizations (Eq.~\ref{dshelm2}, from \cite{eder}).

\section{Charge vs.\ Mass Form Factors}

Neutralinos, of course, interact with the mass distribution of the
nucleus, and for neutralino scattering we should be using mass
rather than charge form factors.  Throughout this paper we have
assumed (as have previous authors, for example \cite{nuclear})
that the mass and charge distributions within nuclei are roughly
proportional.  However, there have been recent attempts to pin
down the nuclear mass distribution using coherent photoproduction
of $\pi^0$ mesons \cite{pionphotoproduction}.  Photoproduction of
pions is in fact an attractive method for studying nuclear mass
distributions; unlike induced hadron reactions it probes the
entire nuclear volume.  In a recent paper on pion photoproduction
\cite{massformfactors} Krusche analyzed the mass form factors of
$^{12}$C, $^{40}$Ca and $^{nat}$Pb using the Helm model for the
form factor.  He found that rms mass radii were slightly smaller
than rms charge radii, though it is still unclear whether these
results are real or merely poorly understood systematic effects in
the data or in the model dependent calculations.  As measurements
of coherent photoproduction of pions from nuclei improve, more
stringent limits of differences between the nuclear mass and
charge densities will be set, including the ability to analyze the
data in a model independent fashion as with electron scattering
data.

\section{Conclusion}

In conclusion, significant differences may exist even at relatively
low momentum transfers between generic Helm-type and more
realistic model independent form factors extracted from elastic
electron scattering data, particularly for large $A$ nuclei. These
new form factors should be utilized to avoid errors in published
limits on neutralino-nucleon cross sections and detection rates in
direct dark matter searches.  The Fourier Bessel and Sum of
Gaussian form factors have the advantage that they are model
independent, analytic, and are derived from electron scattering
data; we have shown that for calculating upper bounds on the
WIMP-proton cross section they are at least as accurate as
modified Helm form factors which have been fit for a specific
nucleus to muonic spectroscopy data. The FB and SOG form factors
have the advantage of existing for the wide array of nuclei being
used in current and future planned dark matter detectors, and
these parameterizations can be used without the need for extensive
fitting to experimental data for different nuclei.  We suggest
that their use is both simpler and more convenient than modified
Helm form factors.  Of course, at present, one still must assume
that the charge distribution in the nucleus mirrors the mass
distribution; however, photo-pion production experiments are
beginning to probe nuclear interiors to give model independent
parameterizations of the nuclear mass density and should provide a
wealth of new data within the next few years.

The model independent analytic form factors in both the Sum of
Gaussians (SOG) and Fourier-Bessel (FB) approaches have been
incorporated into the DarkSUSY \cite{DarkSUSY} code along with a
numerical integration routine to calculate Fourier transforms of
Two-Parameter Fermi distributions; these improvements should
appear in the next major public release of the program and will
replace the need to rely on the simpler Helm form factor.
Parameters in the SOG and FB expansions for the most commonly used
target nuclei are included in the appendix.

\section{Acknowledgements}

G.D. and A.K. would like to thank Creighton University for a
summer research grant which helped support this work. P.G. acknowledges support from the National Science Foundation through grant PHY-0456825.

\newpage

\section{Appendix I: Nuclear Charge Density Parameters for Nuclear Form Factors}

\begin{table}[h]
\begin{center}
\begin{tabular}{|c|l|l|l|l|}
\hline
Nucleus & $^{12}$C & $^{16}$O & $^{28}$Si & $^{30}$Si \\
\hline
rms [fm] & 2.464(12) & 2.737(8) & 3.085(17) & 3.173(25) \\
\hline
$a_1$ & 0.15721e-1 & 0.20238e-1 & 0.33495e-1 & 0.28397e-1 \\
\hline $a_2$ & 0.38897e-1 & 0.44793e-1 & 0.59533e-1  & 0.54163e-1 \\
\hline $a_3$ & 0.37085e-1 & 0.33533e-1& 0.20979e-1 & 0.25167e-1  \\
\hline $a_4$ & 0.14795e-1 & 0.35030e-2 & -0.16900e-1 & -0.12858e-1 \\
\hline $a_5$ & -0.44831e-2 & -0.12293e-1 & -0.14998e-1 & -0.17592e-1 \\
\hline $a_6$ & -0.10057e-1 & -0.10329e-1 & -0.93248e-3 & -0.46722e-2 \\
\hline $a_7$ & -0.68696e-2 & -0.34036e-2 & 0.33266e-2 & 0.24804e-2 \\
\hline $a_8$ & -0.28813e-2 & -0.41627e-3 & 0.59244e-3 & 0.14760e-2 \\
\hline $a_9$ & -0.77229e-3 & -0.94435e-3 & -0.40013e-3 & -0.30168e-3 \\
\hline $a_{10}$ & 0.66908e-4 & -0.25771e-3 & 0.12242e-3 & 0.483464e-4 \\
\hline $a_{11}$ & 0.10636e-3 & 0.23759e-3 & -0.12994e-4 & 0.00000e0 \\
\hline $a_{12}$ & -0.36864e-4 & -0.10603e-3 & -0.92784e-5 & -0.51570e-5 \\
\hline $a_{13}$ & -0.50135e-5 & 0.41480e-4 & 0.72595e-5 & 0.30261e-5 \\
\hline $a_{14}$ & 0.94550e-5 & & -0.42096e-5  &\\
\hline $a_{15}$ & -0.47687e-5 & & &\\
\hline $R$ [fm] & 8.0 & 8.0 & 8.0  & 8.5 \\
\hline
\end{tabular}
\end{center}
\caption{Fourier-Bessel coefficients for $^{12}$C, $^{16}$O,
$^{28}$Si and $^{30}$Si as in \cite{atomic data}, to be used in
~(\ref{eq:FB}).}
\end{table}

\begin{table}[h]
\begin{center}
\begin{tabular}{|c|l|l|l|l|}
\hline
Nucleus & $^{32}$S & $^{40}$Ar & $^{40}$Ca & $^{70}$Ge \\
\hline
rms [fm] & 3.248(4) & 3.423(14) & 3.450(10) & 4.043(2)\\
\hline
$a_1$ & 0.37251e-1 & 0.30451e-1 & 0.44846e-1 & 0.38182e-1\\
\hline $a_2$ & 0.60248e-1 & 0.55337e-1 & 0.61326e-1 & 0.60306e-1\\
\hline $a_3$ & 0.14748e-1 & 0.20203e-1 & -0.16818e-2 & 0.64346e-2 \\
\hline $a_4$ & -0.18352e-1 & -0.16765e-1 & -0.26217e-1 & -0.29427e-1\\
\hline $a_5$ & -0.10347e-1 & -0.13578e-1  & -0.29725e-2 & -0.95888e-2 \\
\hline $a_6$ & 0.30461e-2 & -0.43204e-4 & 0.85534e-2 & 0.87849e-2 \\
\hline $a_7$ & 0.35277e-2 & 0.91988e-3 & 0.35322e-2 & 0.49187e-2 \\
\hline $a_8$ & -0.39834e-4 & -0.41205e-3 & -0.48258e-3 & -0.15189e-2\\
\hline $a_9$ & -0.97177e-4 & 0.11971e-3 & -0.39346e-3 & -0.17385e-2 \\
\hline $a_{10}$ & 0.92279e-4 & -0.19801e-4 & 0.20338e-3 & -0.16794e-3  \\
\hline $a_{11}$ & -0.51931e-4 & -0.43204e-5 & 0.25461e-4 & -0.11746e-3\\
\hline $a_{12}$ & 0.22958e-4 & 0.61205e-5 & -0.17794e-4 & 0.65768e-4\\
\hline $a_{13}$ & -0.86609e-5 & -0.37803e-5 & 0.67394e-5 & -0.30691e-4\\
\hline $a_{14}$ & 0.28879e-5 & 0.18001e-5 & -0.21033e-5 & 0.13051e-5\\
\hline $a_{15}$ & -0.86632e-6 & -0.77407e-6  & & -0.52251e-5 \\
\hline $R$ [fm] & 8.0 & 9.0 & 8.0 & 10.0 \\
\hline
\end{tabular}
\end{center}
\caption{Fourier-Bessel coefficients for $^{32}$S, $^{40}$Ar,
$^{40}$Ca, and $^{70}$Ge as in \cite{atomic data}, to be used in
~(\ref{eq:FB}).}
\end{table}

\begin{table}[h]
\begin{center}
\begin{tabular}{|c|l|l|l|l|}
\hline
Nucleus & $^{72}$Ge & $^{74}$Ge & $^{76}$Ge & $^{208}$Pb \\
\hline
rms [fm] & 4.060(2) & 4.075(2) & 4.081(2) & 5.499(1)\\
\hline
$a_1$ & 0.38083e-1 & 0.37989e-1 & 0.37951e-1 & 0.62732e-1\\
\hline $a_2$ & 0.59342e-1 & 0.58298e-1 & 0.57876e-1 & 0.38542e-1 \\
\hline $a_3$ & 0.47718e-2 & 0.27406e-2 & 0.15303e-2 & -0.55105e-1 \\
\hline $a_4$ & -0.29953e-1 & -0.30666e-1 & -0.31822e-1 & -0.26990e-2\\
\hline $a_5$ & -0.88476e-2 & -0.81505e-2 & -0.76875e-2 & 0.31016e-1\\
\hline $a_6$ & 0.96205e-2 & 0.10231e-1 & 0.11237e-1 & -0.99486e-2\\
\hline $a_7$ & 0.47901e-2 & 0.49382e-2 & 0.50780e-2 & -0.93012e-2\\
\hline $a_8$ & -0.16869e-2 & -0.16270e-2 & -0.17293e-2 & 0.76653e-2\\
\hline $a_9$ & -0.15406e-2 & -0.13937e-2 & -0.15523e-2 & 0.20886e-2\\
\hline $a_{10}$ & -0.97230e-4 & 0.15376e-3 & 0.72439e-4 & -0.17840e-2\\
\hline $a_{11}$ & -0.47640e-4 & 0.14396e-3 & 0.16560e-3& 0.74876e-4 \\
\hline $a_{12}$ & -0.15669e-5 & -0.73075e-4 & -0.86631e-4 & 0.32278e-3\\
\hline $a_{13}$ & 0.67076e-5 & 0.31998e-4 & 0.39159e-4 & -0.11353e-3\\
\hline $a_{14}$ & -0.44500e-5 & -0.12822e-4 & -0.16259e-4 &\\
\hline $a_{15}$ & 0.22158e-5 & 0.48406e-5 & 0.63681e-5 &\\
\hline $R$ [fm] & 10.0 & 10.0  & 10.0 &  11.0\\
\hline
\end{tabular}
\end{center}
\caption{Fourier-Bessel coefficients for $^{72}$Ge, $^{74}$Ge,
$^{76}$Ge, and $^{208}$Pb as in \cite{atomic data}, to be used in
(\ref{eq:FB}).}
\end{table}

\begin{table}[h]
\begin{center}
\begin{tabular}{|c|c|c|c|c|c|c|}
\hline
& \multicolumn{2}{|c|}{$^{12}$C} & \multicolumn{2}{|c|}{$^{16}$O} & \multicolumn{2}{|c|}{$^{28}$Si} \\
\hline
rms [fm] & \multicolumn{2}{|c|}{2.469(6)} & \multicolumn{2}{|c|}{2.711} & \multicolumn{2}{|c|}{3.121}  \\
\hline $i$ & $R_i$ & $Q_i$ & $R_i$ & $Q_i$ & $R_i$ & $Q_i$ \\
\hline 1 & 0.0 & 0.016690 & 0.4 & 0.057056 & 0.4 & 0.033149 \\
\hline 2 & 0.4 & 0.050325 & 1.1 & 0.195701 & 1.0 & 0.106452 \\
\hline 3 & 1.0 & 0.138621 & 1.9 & 0.311188 & 1.9 & 0.206866 \\
\hline 4 & 1.3 & 0.180515 & 2.2 & 0.224321 & 2.4 & 0.286391  \\
\hline 5 & 1.7 & 0.219097 & 2.7 & 0.059946 & 3.2 & 0.250448 \\
\hline 6 & 2.3 & 0.278416 & 3.3 & 0.135714 & 3.6 & 0.056944 \\
\hline 7 & 2.7 & 0.058779 & 4.1 & 0.000024 & 4.1 & 0.016829 \\
\hline 8 & 3.5 & 0.057817 & 4.6 & 0.013961 & 4.6 & 0.039630 \\
\hline 9 & 4.3 & 0.007739 & 5.3 & 0.000007 & 5.1 & 0.000002 \\
\hline 10 & 5.4 & 0.02001 & 5.6 & 0.000002 & 5.5 & 0.000938 \\
\hline 11 & 6.7 & 0.00007 & 5.9 & 0.002096 & 6.0 & 0.000002\\
\hline 12 & & & 6.4 & 0.000002 & 6.9 & 0.002366 \\
\hline RP [fm] & \multicolumn{2}{|c|}{1.20} & \multicolumn{2}{|c|}{1.30} & \multicolumn{2}{|c|}{1.30} \\
\hline
\end{tabular}
\end{center}
\caption{Sum of Gaussian coefficients for $^{12}$C, $^{16}$O, and
$^{28}$Si as in \cite{atomic data}, to be used in ~(\ref{eq:SOG}).
RP, the rms radius of the Gaussians, is related to $\gamma$ by RP
= $\gamma \sqrt{3/2}$.}
\end{table}

\begin{table}[h]
\begin{center}
\begin{tabular}{|c|c|c|c|c|c|c|}
\hline
& \multicolumn{2}{|c|}{$^{32}$S} & \multicolumn{2}{|c|}{$^{40}$Ca} & \multicolumn{2}{|c|}{$^{208}$Pb} \\
\hline
rms [fm] & \multicolumn{2}{|c|}{3.258} & \multicolumn{2}{|c|}{3.480(3)} & \multicolumn{2}{|c|}{5.503(2)}  \\
\hline $i$ & $R_i$ & $Q_i$ & $R_i$ & $Q_i$ & $R_i$ & $Q_i$ \\
\hline 1 & 0.4 & 0.045356 & 0.4 & 0.042870 & 0.1 & 0.003845\\
\hline 2 & 1.1 & 0.067478 & 1.2 & 0.056020 & 0.7 & 0.009724\\
\hline 3 & 1.7 & 0.172560 & 1.8 & 0.167853 & 1.6 & 0.033093\\
\hline 4 & 2.5 & 0.324870 & 2.7 & 0.317962 & 2.1 & 0.000120\\
\hline 5 & 3.2 & 0.254889 & 3.2 & 0.155450 & 2.7 & 0.083107\\
\hline 6 & 4.0 & 0.101799 & 3.6 & 0.161897 & 3.5 & 0.080869\\
\hline 7 & 4.6 & 0.022166 & 4.3 & 0.053763 & 4.2 & 0.139957\\
\hline 8 & 5.0 & 0.002081 & 4.6 & 0.032612 & 5.1 & 0.260892\\
\hline 9 & 5.5 & 0.005616 & 5.4 & 0.004803 & 6.0 & 0.336013\\
\hline 10 & 6.3 & 0.000020 & 6.3 & 0.004541 & 6.6 & 0.033637\\
\hline 11 & 7.3 & 0.000020 & 6.6 & 0.000015 & 7.6 & 0.018729\\
\hline 12 & 7.7 & 0.003219 & 8.1 & 0.002218 & 8.7 & 0.000020\\
\hline RP [fm] & \multicolumn{2}{|c|}{1.35} & \multicolumn{2}{|c|}{1.45}  & \multicolumn{2}{|c|}{1.70}  \\
\hline
\end{tabular}
\end{center}
\caption{Sum of Gaussian coefficients for $^{32}$S, $^{40}$Ca, and
$^{208}$Pb as in \cite{atomic data}, to be used in
~(\ref{eq:SOG}).  RP, the rms radius of the Gaussians, is related
to $\gamma$ by RP = $\gamma \sqrt{3/2}$.}
\end{table}

\begin{table}[h]
\begin{center}
\begin{tabular}{|c|c|c|c|}
\hline
Nucleus & c [fm] & $\left<r^2\right>^{1/2}$ [fm] & a [fm] \\
\hline $^{23}$Na & 2.9393 & 2.994 & 0.523 \\
\hline $^{127}$I & 5.5931 & 4.749 & 0.523 \\
\hline $^{129}$Xe & 5.6315 & 4.776 & 0.523 \\
\hline $^{131}$Xe & 5.6384 & 4.781 & 0.523 \\
\hline $^{132}$Xe & 5.6460 & 4.787 & 0.523 \\
\hline $^{134}$Xe & 5.6539 & 4.792 & 0.523 \\
\hline $^{184}$W & 6.51(7) & 5.42(7) & 0.535(36) \\
\hline $^{186}$W & 6.58(3) & 5.40(4) & 0.480(23) \\
\hline
\end{tabular}
\end{center}
\caption{Two-Parameter Fermi coefficients for $^{23}$Na,
$^{127}$I, and $^{129-134}$Xe as in \cite{muon data new} and for
$^{184,186}$W as in \cite{atomic data}, to be used in
~(\ref{eq:2ptFermi}). $\left<r^2\right>^{1/2}$, the rms value of
the charge radius calculated using the Two-Parameter Fermi
distribution with $t=2.30$ fm for $^{23}$Na, $^{127}$I, and
$^{129-134}$Xe, and $c$ and $a$ as given above for $^{184,186}$W,
is also given.}
\end{table}


\begin{thebibliography}{99}

\bibitem{1} Bergstr\"{o}m L and Gondolo P 1996 \textit{Astrop. Phys.} \textbf{5} 263

\bibitem{2} Bergstr\"{o}m L, Edsj\"{o} J and Gondolo P 1997 \textit{Phys. Rev. D} \textbf{55} 1765

\bibitem{3} Bergstr\"{o}m L, Edsj\"{o} J and Ullio P 1998 \textit{Phys. Rev. D} \textbf{58}
083507

\bibitem{4} Bergstr\"{o}m L,  Edsj\"{o} J and Ullio P 1999,\textit{Astrophys. J.} \textbf{526} 215

\bibitem{5} Baltz E A and Edsj\"{o} J 1999 \textit{Phys. Rev. D} \textbf{59} 023511

\bibitem{6} Edsj\"{o} J and Gondolo P 1995 \textit{Phys. Lett. B} \textbf{357} 595

\bibitem{7} Edsj\"{o} J, PhD Thesis (\textit{Preprint}
hep-ph/9704384)

\bibitem{8} Bergstr\"{o}m L, Damour T, Edsj\"{o} J, Krauss L and Ullio P 1999 \textit{J. High Energy Phys.} JHEP\textbf{9908}
010

\bibitem{nuclear} Engel J, Pittel S, and Vogel P 1992 \textit{Int. Journal
of Mod. Phys. E} \textbf{1} 1-37

\bibitem{smith} Lewin J D and Smith P F 1996 \textit{Astropart. Phys.} \textbf{6} 87-112.

\bibitem{gondolo96} Gondolo P 1997, in {\it Dark Matter, Quantum Measurements and Experimental Gravitation}, XXXI Rencontres de Moriond, Leas Arcs 1996, ed. J. Tr\^an Thanh V\^an (Editions Fronti\`ere, 1997).

\bibitem{compatibility with dama} Gondolo P and Gelmini G 2005
\textit{Phys. Rev. D} \textbf{71} 123520

\bibitem{Uberall} Uberall P 1971 \textit{Electron Scattering from Complex Nuclei} (New York: Academic
Press)

\bibitem{DarkSUSY} Gondolo P, Edsj\"{o} J, Ullio P, Bergstr\"{o}m L, Schelke M, and Baltz E A 2004
\textit{J. Cosmol. Astropart. Phys.} \textbf{7}

\bibitem{eder} Eder G 1968 \textit{Nuclear Forces} (MIT Press)

\bibitem{jkg} Jungman G, Kamionkowski M, and Griest K 1996 \textit{Phys. Rep.} \textbf{267} 195

\bibitem{engel} Engel J 1991 \textit{Phys. Lett.} \textbf{B 264} 114

\bibitem{Helm} Helm R 1956 \textit{Phys. Rev.} \textbf{104} 1466

\bibitem{Hofstadter} Hofstadter R et al. 1977 \textit{Phys. Rev.
Lett.} \textbf{38} 152

\bibitem{atomic data} De Vries H et al. 1987 \textit{Atomic Data and Nuclear Data} \textbf{36} 495-529

\bibitem{atomic data old} De Jager et al. 1974 \textit{Atomic Data and Nuclear
Data Tables} \textbf{14} 479

\bibitem{muon data old} Engfer R et al. 1974 \textit{Atomic Data and Nuclear
Data Tables} \textbf{14} 509-597

\bibitem{muon data new} Fricke G et al. 1995 \textit{Atomic Data and Nuclear
Data Tables} \textbf{60} 177-285

\bibitem{barret moments} Barrett R 1970 \textit{Phys. Rev. Lett. B}
\textbf{33} 388

\bibitem{Frauenfelder} Frauenfelder H and Henley E 1991
\textit{Subatomic Physics} 2nd Ed. (New York: Prentice Hall)

\bibitem{Sick1974} Sick I 1974 \textit{Nucl. Phys. A}
\textbf{218} 509-541

\bibitem{dreher} Dreher et al. 1974 \textit{Nucl. Phys. A} \textbf{235} 219-248.

\bibitem{anni} Anni R, Co' G and Pellegrino P 1995 \textit{Nucl.Phys. A} \textbf{584} 35-59

\bibitem{DAMA} Bernabei R et al. [DAMA col.] 2000 \textit{Phys. Lett. B} {\textbf 480} 23

\bibitem{ZEPLIN} Alner G J et al. [UK Dark Matter Collaboration] 2005 \textit{Astropart. Phys.}
\textbf{23} 444-462

\bibitem{XENON} Aprile E et al. 2005 \textit{New Astron. Rev.}
\textbf{49} 289-295

\bibitem{CDMS} Akerib D S et al. 2006 \textit{Phys. Rev. Lett.} \textbf{96} 011302

\bibitem{EDELWEISS} Luca M [EDELWEISS Collaboration] (Preprint astro-ph/0605496)

\bibitem{DRIFT} Alner G J et al. 2005 \textit{Nucl. Instrum. Meth. A}
\textbf{555} 173

\bibitem{CRESST} Angloher G et. al. 2005, \textit{Astropar. Phys.} \textbf{23} 325-339

\bibitem{CryoArray} Schnee R, Akerib D, and Gaitskell R 2003 \textit{Nucl. Phys. Proc. Suppl.}
\textbf{124} 233-236

\bibitem{Gaitskell} Gaitskell R 2004 \textit{Annu. Rev. Nucl. Part.
Sci.} \textbf{54} 315-359

\bibitem{sagittarius stream} Freese K, Gondolo P, and Newberg H
2005 \textit{Phys. Rev. D} \textbf{71} 043516

\bibitem{CDMS-Soudan constraints} Akerib DS et al. 2004 \textit{Phys
Rev Lett.} \textbf{93} 211301

\bibitem{pionphotoproduction} Krusche B et al. 2002 \textit{Phys. Rev. Lett.
B} \textbf{526} 287

\bibitem{massformfactors} Krusche B 2005 \textit{Eur. Phys. J. A}
\textbf{26} 7-18


\end{thebibliography}
\end{document}